\documentclass[letterpaper]{article}
\usepackage{natbib,alifeconf}
\usepackage{amsmath}
\usepackage{booktabs}

\title{Self-Regulation through Communication\\in Evolved Neural Agents}

\author{Josh Nunley$^{1}$ \\
\mbox{}\\
$^1$Indiana University, Bloomington, IN, USA \\
joshnunl@iu.edu}

\begin{document}

\maketitle

\begin{abstract}
Communication is typically understood as \textit{indication}: signals that transfer information from sender to receiver. We present a minimal predator avoidance task in which pairs of evolved CTRNN agents use communication for robust survival, and in which agents hear their own vocalizations, as in natural systems. Across 112 perfect-fitness agents from over 2{,}000 evolutionary runs, three dominant strategies emerge (accounting for 81\% of agents): \textbf{safety calling} (39\%), where agents signal from safe cover; \textbf{alarm indication} (22\%), where agents vocalize when a threat is present without relying on self-hearing; and \textbf{self-regulatory calling} (20\%), where agents depend on hearing their own call to sustain escape behavior. Self-hearing dependency is common among agents that call during an active threat (47\%), but rare among agents that call only after reaching safe cover (10\%; $p < 10^{-4}$). This pattern is consistent with a difference in causal order: safety callers act then communicate, while self-regulatory callers communicate in order to act. Removing self-hearing selectively impairs self-regulatory callers (fitness 0.40) while safety callers remain functional (0.90; $p < 10^{-9}$). These results show that communication can evolve to serve the caller's own behavioral regulation, not just information transfer to others.
\end{abstract}

\section{Introduction}

Alarm calls present a puzzle for evolutionary biology: why would an organism advertise its presence to predators by vocalizing? The classic demonstration comes from vervet monkeys, which produce acoustically distinct alarm calls for different predator types. Receivers respond with predator-appropriate evasive action: hiding in bushes for eagles, climbing trees for leopards \citep{seyfarth1980monkey}. This finding became a landmark in animal communication research: alarm calls can encode specific information about threat type, and receivers respond adaptively to that information. This is a pattern widely interpreted as evidence for \textit{indication}, the view that communication functions to transfer information from sender to receiver \citep{hauser1996evolution}.

However, this indication-centered framing may underestimate the functional diversity of alarm communication. Several observations suggest that vocalizations may serve functions beyond information transfer. First, some alarm calls are still produced when callers are solitary or when the social audience is reduced \citep{townsend2012flexible}. Second, the act of calling often appears tightly coupled to the caller's own behavioral response, raising questions about whether the call causes, or merely accompanies, appropriate action \citep{owings1977snake, owings1998animal}. Third, signals may function to manipulate receivers' behavior \citep{dawkins1978animal} or influence their states \citep{owren2010redefining} rather than to provide accurate information. These observations are difficult to explain if communication serves only to indicate information to others.

We propose that alarm calls may additionally serve a \textit{self-regulatory} function: the act of producing a signal may itself sustain or modulate the signaler's behavioral response through self-hearing. In vocal learning systems, auditory feedback is known to affect the production and maintenance of vocalizations \citep{konishi1965effects, brainard2000auditory, eliades2008neural}. The Lombard effect shows that vocal organisms adjust amplitude and other vocal features in noisy conditions \citep{zollinger2011lombard}. Yet prior work has focused on how self-hearing affects the \textit{production} of vocalizations, rather than on whether it affects their \textit{behavioral function}. In other words, does a caller's own signal help regulate its actions?

Computational models of evolved communication provide a natural testbed for this hypothesis. Within the minimal cognition tradition \citep{beer2003dynamics}, Williams, Beer, and colleagues evolved pairs of CTRNN agents to perform referential communication tasks \citep{williams2008evolving}, an approach since extended to conceptual \citep{yao2023go} and displaced communication \citep{bernard2023evolution}. At larger scale, deep multi-agent reinforcement learning has investigated emergent communication \citep{foerster2016learning, lazaridou2020emergent}. However, these models typically treat communication as a unidirectional channel from sender to receiver, omitting a basic feature of biological vocalization: \textit{organisms hear their own calls}. This self-hearing channel creates the possibility for communication to serve self-regulatory functions. Without it, there is no feedback pathway through which a vocalization could influence the caller's own behavior. Furthermore, most studies analyze one or a few evolved solutions, making it difficult to characterize the diversity of strategies that a given task makes possible.

To address both gaps, we developed a minimal predator avoidance task that abstracts the core structure of the vervet monkey paradigm: a predator targets a specific cover type, one agent receives warning of which type is safe, and the partner must rely on communication to reach the correct cover reliably. Critically, each agent hears its own vocalizations, as in natural systems. From over 2{,}000 independent evolutionary runs, we selected 112 agents that achieved perfect fitness, enabling population-level analysis of evolved communication strategies. Three functionally distinct strategies emerge with comparable frequency: \textit{safety calling}, where agents signal from safe cover to guide their partner; \textit{alarm indication}, where agents signal threat presence; and \textit{self-regulatory calling}, where agents depend on hearing their own vocalization to sustain escape behavior. Self-hearing ablation reveals a difference in causal order between these strategies: safety callers act first and then communicate, while self-regulatory callers communicate in order to act. These results show that even a single minimal task can give rise to qualitatively different relationships between communication and behavior.

\section{Methods}

\subsection{Task Environment}

\begin{figure}[t]
    \centering
    \includegraphics[width=0.8\linewidth]{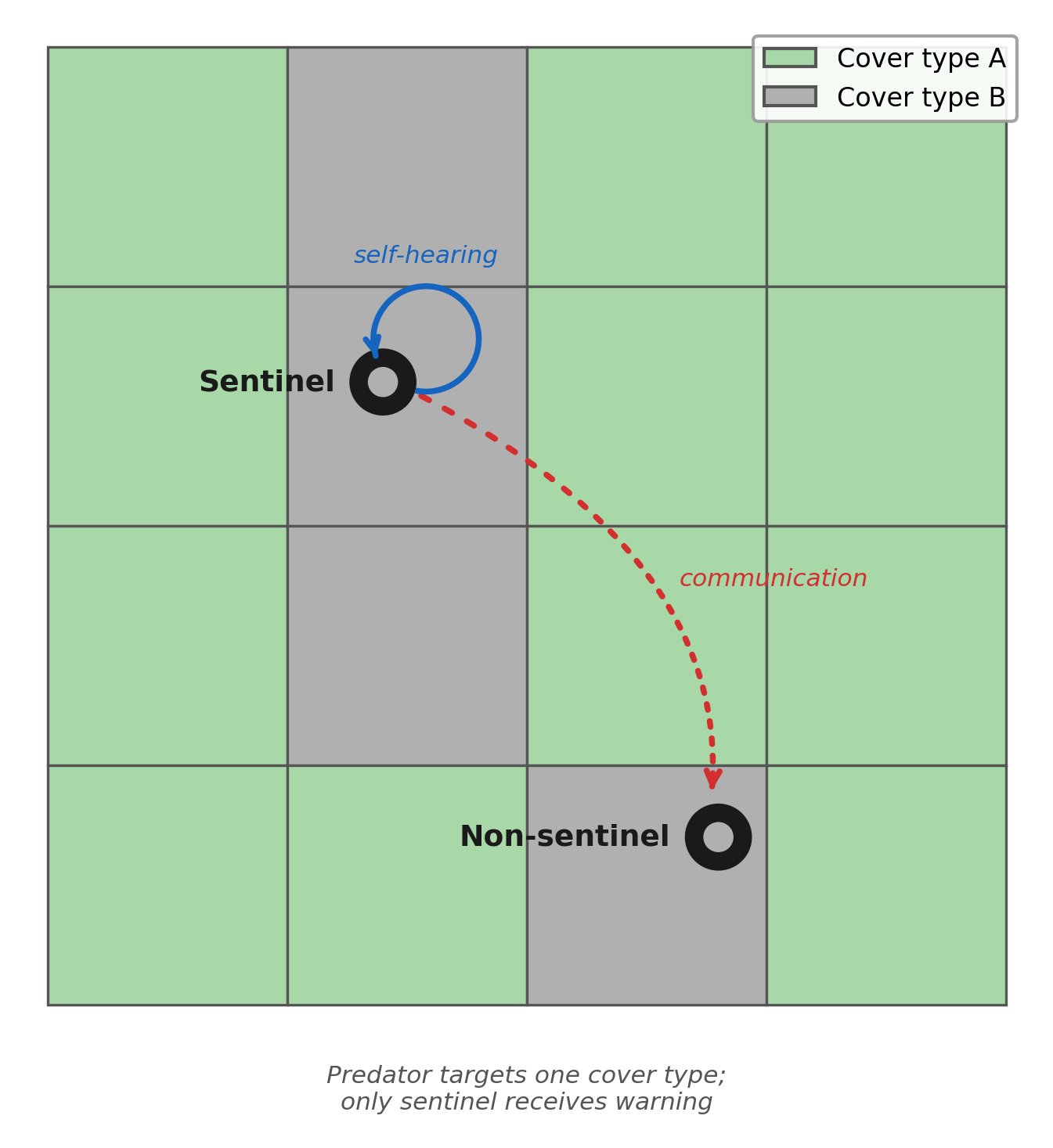}
    \caption{Task environment snapshot. The 4$\times$4 grid overlays the continuous space, with each cell randomly assigned to one of two cover types (shown as different colors). Agents (circles) must reach a cell of the correct cover type before each predator attack. Only the sentinel agent receives direct warning of which cover type will be safe.}
    \label{fig:environment}
\end{figure}

We designed a grid-based environment in which two agents must survive periodic predator attacks by reaching appropriate cover (Figure~\ref{fig:environment}). A 10$\times$10 continuous space (toroidal) is overlaid with a 4$\times$4 grid; each cell is randomly assigned one of two cover types. Periodically (every 150 timesteps; $dt = 0.1$), a predator attacks; one cover type provides safety and agents in cells of that type survive, while agents in the other type are eliminated. Each trial lasts 750 timesteps (5 attacks).

The key constraint is \textbf{asymmetric information}: only one randomly designated agent (the ``sentinel'') receives a warning indicating which cover type will be safe. The non-sentinel must rely on communication to select appropriate cover reliably.

\subsection{Agent Architecture}

Each agent is controlled by a continuous-time recurrent neural network (CTRNN) with 5 fully-connected hidden neurons \citep{beer1995dynamical}. The network dynamics follow:
\begin{equation}
    \tau_i \frac{dy_i}{dt} = -y_i + \sum_j w_{ji} \sigma(y_j + \theta_j) + \sum_k w_{ki}^{\text{in}} I_k
\end{equation}
where $y_i$ is the activation of neuron $i$, $\tau_i$ is its time constant, $w_{ji}$ are recurrent weights, $\theta_j$ are biases, $\sigma(x) = 1/(1+e^{-x})$ is the logistic sigmoid, and $I_k$ are external inputs with weights $w_{ki}^{\text{in}}$. Weights, biases, and time constants are inherited from the evolved genotype without hand-tuned bounds or post-hoc adjustment.

The network receives 5 inputs: a 2-dimensional one-hot cover sensor, a 2-dimensional one-hot predator warning (active only for the sentinel; zeros for the non-sentinel), and 1 communication channel (summing the partner's output and the agent's own previous output). It produces 3 sigmoid outputs: two movement controls and one communication output, which is binarized by rounding for transmission. This gives 83 evolved parameters (input weights and biases, recurrent weights and biases, time constants, and output weights and biases).

Each agent's communication output at timestep $t$ is included in its own communication input at timestep $t+1$, with no attenuation. This \textbf{self-hearing} channel mirrors the basic fact that biological organisms hear their own vocalizations. Both agents in a pair share identical network weights (clonal population) but differ in their designated role each attack cycle.

\subsection{Evolution}

We evolved agent networks using PyGAD \citep{pygad} with steady-state selection, single-point crossover, and uniform additive mutation (5\% of genes per offspring; perturbation drawn from $U(-1, +1)$; population 30; 5 parents; 1 elite preserved; 150 generations). Each genotype is evaluated over 20 trials with randomized configurations, using a fitness function $F = 0.9 \times \min(f) + 0.1 \times \text{mean}(f)$ where $f$ is the survival proportion per trial. This weighting emphasizes worst-case robustness; achieving $F = 1.0$ requires perfect survival across all 20 evaluation environments.

We conducted over 2{,}000 independent evolutionary runs with self-hearing enabled. From these, we selected the 112 agents that achieved perfect fitness ($F = 1.0$), representing approximately 5\% of total runs. Perfect fitness is a stringent criterion: the remaining $\sim$95\% of runs produced agents with high but imperfect survival. All subsequent analyses are conducted on these perfect-fitness agents.

\subsection{Systematic Input-Output Profiling}
\label{sec:profiling}

To classify evolved agents, we measure each agent's communication and movement responses under controlled input conditions. For each agent, we fix the sensory inputs and run the network for 100 timesteps. We define three semantic conditions: \textbf{SAFE} (cover matches predator target), \textbf{UNSAFE} (cover does not match), and \textbf{NO\_PREDATOR} (no predator warning). Since there are two cover types and three predator states, this yields six input combinations; we average across cover types to obtain three condition-level measurements. Each agent is profiled with self-hearing enabled and disabled.

\textbf{Self-hearing dependency:} For each agent, we compare total communication output and movement discrimination (unsafe minus safe movement) with and without self-hearing. An agent is classified as \textit{self-hearing dependent} if removing self-hearing causes a large change in total communication (difference $> 1.0$) or eliminates differential movement responses (unsafe--safe movement difference $< 1.0$).

\subsection{Transition Profiling}
\label{sec:transition}

We also measure how agents respond to sudden transitions between SAFE and UNSAFE conditions. For each test, we run 50 burn-in timesteps with one condition, then switch the cover input and measure for 50 additional timesteps. We test both SAFE$\to$UNSAFE and UNSAFE$\to$SAFE transitions.

\subsection{Two-Agent Ecological Ablation}
\label{sec:ecological}

To validate that controlled classifications predict behavior in the full ecological setting, we used ablation studies. In these experiments, specific communication channels are selectively disabled to test their causal contribution. We tested all 112 agents in the complete two-agent task under four communication conditions: \textbf{Full} (both channels enabled), \textbf{NoSelf} (self-hearing disabled), \textbf{NoSocial} (social hearing disabled), and \textbf{NoComm} (both disabled). Each condition was evaluated over 20 random seeds using the same per-trial survival scoring as evolution. The ecological analyses below use mean fitness across seeds.

\subsection{Statistics}

Group differences were assessed with nonparametric tests (Kruskal--Wallis for multi-category, Mann--Whitney for pairwise comparisons, Fisher exact for proportions), using each evolved agent as the unit of analysis.

\section{Results}

\subsection{Three Communication Strategies Emerge}

\begin{figure}[t]
    \centering
    \includegraphics[width=0.85\linewidth]{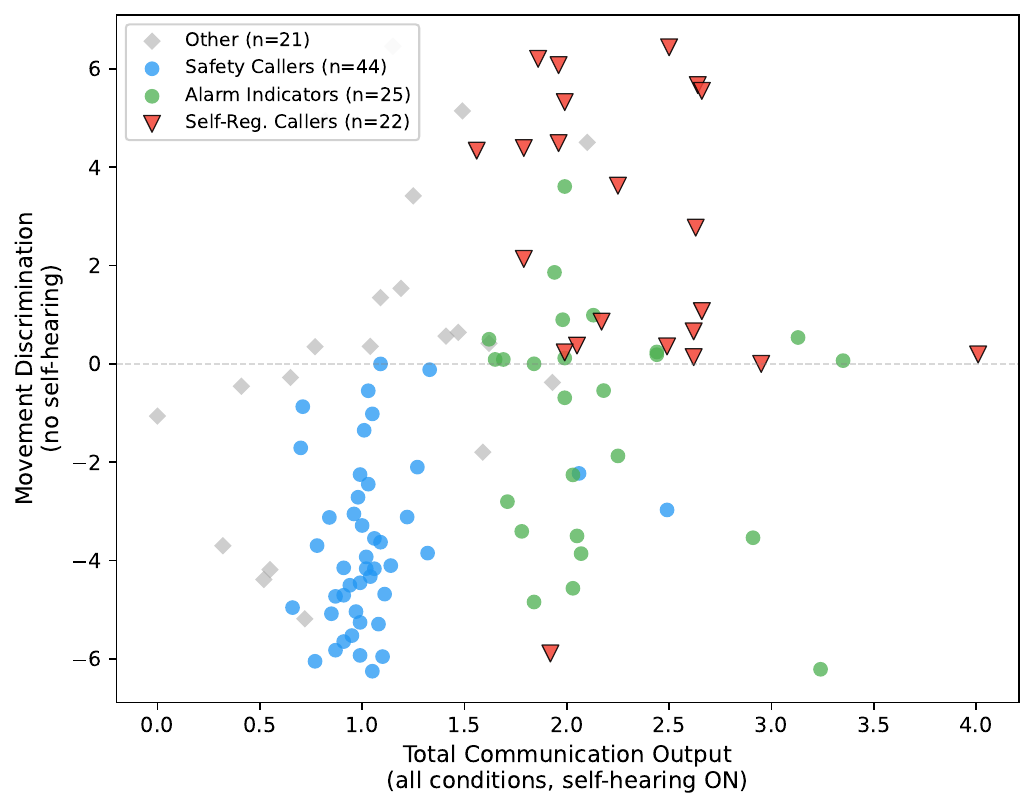}
    \caption{Agent behavioral landscape. Each point represents one of 112 evolved agents, positioned by total communication output (x-axis) and movement discrimination without self-hearing (y-axis). Safety callers (blue circles) cluster at lower communication levels. Alarm indicators (green circles) and self-regulatory callers (red triangles, black-edged) span higher communication levels; self-regulatory agents show reduced movement discrimination, indicating they cannot sustain appropriate movement without self-hearing. Grey diamonds show agents with other communication patterns.}
    \label{fig:classification}
\end{figure}

\begin{figure*}[t]
    \centering
    \includegraphics[width=\linewidth]{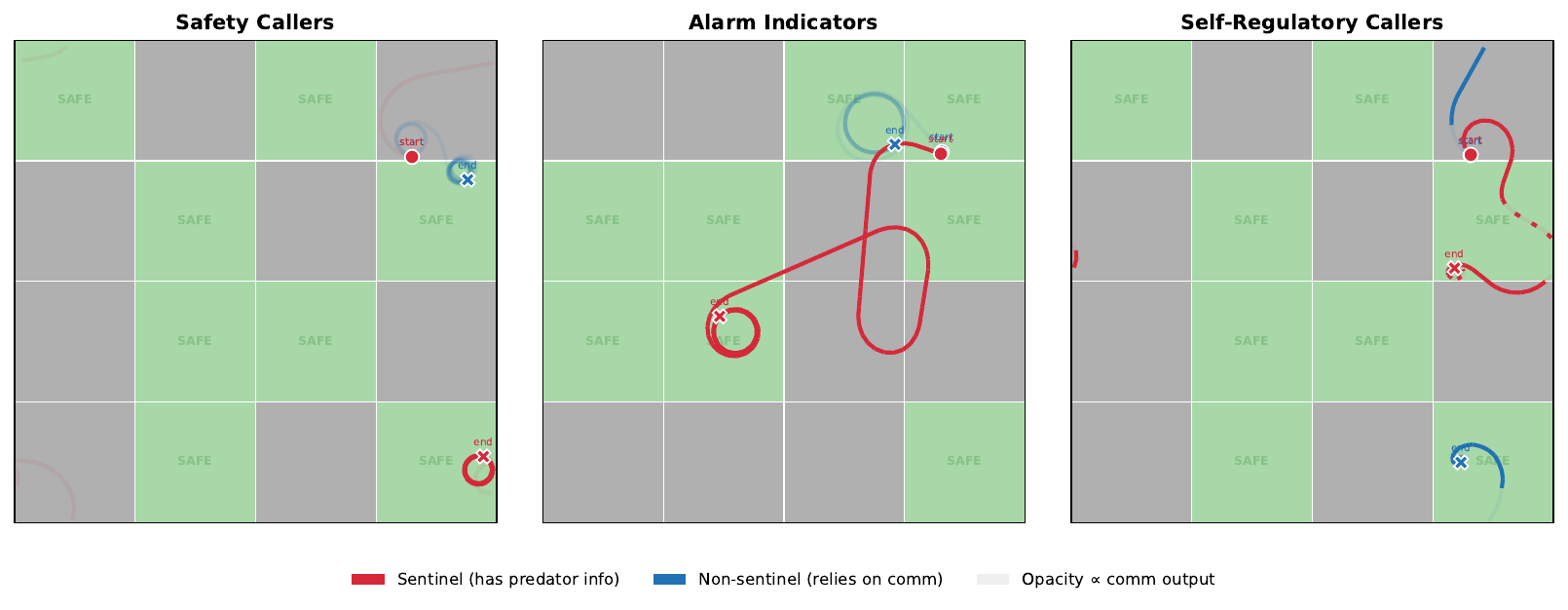}
    \caption{Representative movement traces for a single predator attack cycle, one agent from each strategy. The sentinel (red; has predator information) and non-sentinel (blue; must rely on communication) are shown moving across the cover grid; filled circles mark each agent's starting position and $\times$ marks the position at the moment of the predator attack. Trajectory opacity encodes communication output at each timestep (transparent $=$ silent, opaque $=$ calling). Cells labeled \textsc{safe} indicate the correct cover type for this attack. \textit{Safety caller} (left): the sentinel produces little to no communication during escape, brightening only upon reaching safe cover. Here, communication is a consequence of safety achieved, not a means of achieving it. \textit{Alarm indicator} (center): the sentinel signals immediately at threat onset and maintains high output throughout; the non-sentinel's trajectory is largely transparent, showing it does not depend on its own vocalization to sustain escape. \textit{Self-regulatory caller} (right): both agents maintain high communication output throughout the episode. As the ablation results below show, the sentinel's own calling helps sustain its escape. The non-sentinel, sharing the same neural weights, likewise vocalizes persistently, creating a mutually reinforcing signaling loop.}
    \label{fig:movement_traces}
\end{figure*}

Despite identical architectures and task constraints, the 112 evolved agents display qualitatively diverse communication strategies. Using the systematic profiling described in Methods, we classify agents along two dimensions: \textit{when} they communicate (in safe conditions only, or whenever a predator is present) and \textit{whether they depend on self-hearing} (whether removing self-hearing disrupts their behavior). Together, these dimensions yield three functionally distinct strategies (Figure~\ref{fig:classification}, Figure~\ref{fig:movement_traces}, Table~\ref{tab:dependency}):

\begin{enumerate}
    \item \textbf{Safety callers} (44 agents, 39\%): These agents vocalize specifically when they have reached safe cover during a threat. Removing self-hearing does not disrupt their behavior; their vocalization follows as a consequence of the agent's state change upon reaching safety, and the partner uses this signal to locate appropriate cover.
    \item \textbf{Alarm indicators} (25 agents, 22\%): These agents vocalize whenever a predator is present, regardless of their own safety status. Like safety callers, they function without self-hearing; they signal threat, and the partner uses this signal to guide its escape.
    \item \textbf{Self-regulatory callers} (22 agents, 20\%): These agents also vocalize during threat, but unlike alarm indicators, they \textit{depend on hearing their own call}. As the ablation results below show, their vocalization serves a dual function: indicating threat to the partner \textit{and} sustaining the caller's own escape behavior through self-hearing.
\end{enumerate}

\begin{table}[ht]
\centering
\small
\begin{tabular}{lccc}
\toprule
Vocalization timing & $n$ & \% SH-Dep. & \% Not Dep. \\
\midrule
From safety & 49 & 10.2\% & 89.8\% \\
During threat & 47 & 46.8\% & 53.2\% \\
\bottomrule
\end{tabular}
\caption{Self-hearing dependency by communication timing. Among agents that vocalize during threat (alarm indicators and self-regulatory callers combined), nearly half depend on self-hearing, approximately 4.5$\times$ the rate among agents that vocalize from safety.}
\label{tab:dependency}
\end{table}

Self-hearing dependency is asymmetric: nearly half of agents that vocalize during threat depend on hearing their own call, compared with only one in ten agents that vocalize from safety ($p < 10^{-4}$, Fisher exact test). This asymmetry is explained in the following section.

The remaining 21 agents include 5 self-hearing dependent safety-calling agents, 4 agents that communicate only when in danger, and 12 agents with no differential \textit{steady-state} communication. These 12 agents still show reduced survival when communication is removed in the full task. Four of them respond to transitions (2 when safety is reached, 2 at both transitions), suggesting they communicate transiently rather than in steady state; the remaining 8 show no response to either profiling method. We focus subsequent analysis on the three dominant strategies, which account for 91 of 112 agents (81\%).

\subsection{Causal Direction Explains the Asymmetry}

The concentration of self-hearing dependency among agents that call during threat reflects a difference in the \textit{causal direction} between action and communication across the three strategies.

\textbf{Safety callers act first, then communicate.} A safety caller moves to appropriate cover and, once there, begins vocalizing. Self-hearing is unnecessary because the vocalization is a \textit{consequence} of having reached the correct behavioral state, not a driver of it.

\textbf{Self-regulatory callers communicate in order to act.} When a self-regulatory caller detects a predator warning, it begins to vocalize. This vocalization feeds back through the self-hearing channel, sustaining the neural dynamics driving escape behavior. When self-hearing is removed, these agents can no longer maintain different movement responses to safe versus unsafe conditions.

\textbf{Alarm indicators} represent a distinct solution: agents whose recurrent dynamics sustain escape behavior independently, without requiring either the completion of escape (as in safety callers) or self-hearing. They vocalize during threat but do not depend on hearing their own call.

\begin{figure*}[t]
    \centering
    \includegraphics[width=\linewidth]{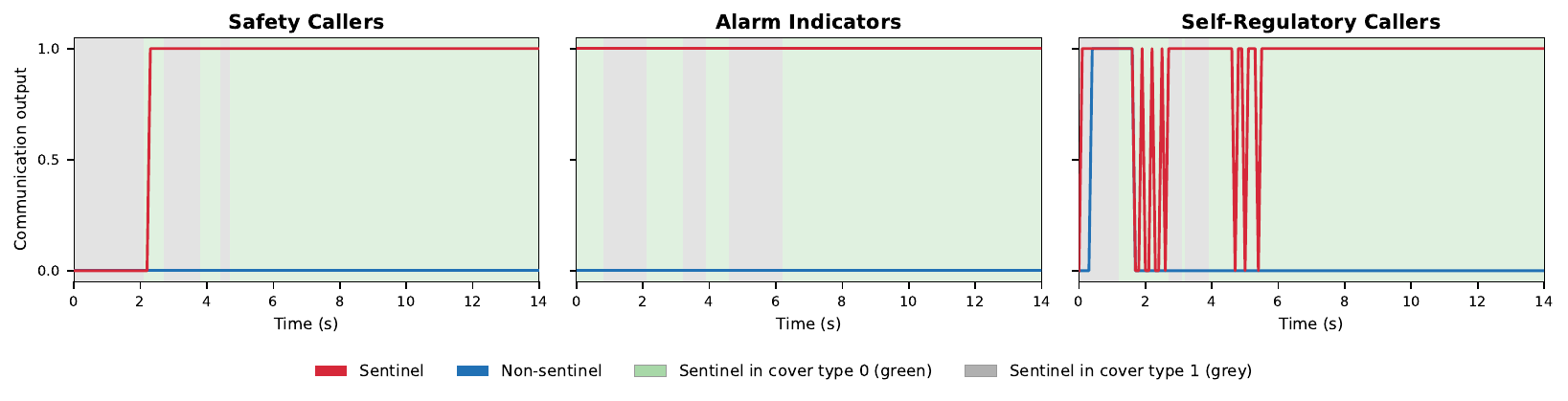}
    \caption{Communication output over a single predator attack cycle for a representative agent from each strategy. Red = sentinel; blue = non-sentinel. Background shading shows the cover type the sentinel currently occupies (green = cover type 0; grey = cover type 1), making it easy to see when the sentinel reaches safe cover. \textit{Safety callers} (left): the sentinel is silent until it reaches safe cover. Communication then rises sharply, so the background color shift and communication onset occur together. The non-sentinel remains largely silent throughout. \textit{Alarm indicators} (center): the sentinel signals immediately at threat onset and maintains high output regardless of cover type; the non-sentinel produces little communication, consistent with its independence from self-hearing. \textit{Self-regulatory callers} (right): both agents sustain high communication output for the full duration of the threat, consistent with a mutually reinforcing signaling loop that contributes to escape behavior.}
    \label{fig:traces}
\end{figure*}

Figure~\ref{fig:traces} illustrates these differences. The safety caller (left) communicates only when safe, with communication tracking the agent's safety state. The alarm indicator (middle) maintains communication regardless of safety state. The self-regulatory caller (right) shows tightly coupled communication and movement, with both channels switching together at the transition point.

Transition profiling shows the same pattern. Among all agents that vocalize from safety ($n=49$), the most common transition response is to begin communicating when safety is reached (15/49). Among all agents that vocalize during threat ($n=47$), most respond to \textit{both} transition directions (26/47), consistent with general threat-responsive communication. Agents with no differential steady-state communication are mostly non-responsive to transitions as well (11/16), suggesting that they communicate little in these controlled tests.

\subsection{Causal Test: Self-Hearing Ablation in the Ecological Task}
\label{sec:ecological_results}

If self-regulatory callers communicate \textit{in order to} act, while safety callers act independently of their own communication, then removing self-hearing should most severely impair self-regulatory callers. Safety callers should show the least disruption. Strategy categories were defined using single-agent controlled profiling; the ecological fitness results reported here test whether those categories predict behavior in the full task.

\begin{figure}[t]
    \centering
    \includegraphics[width=\linewidth]{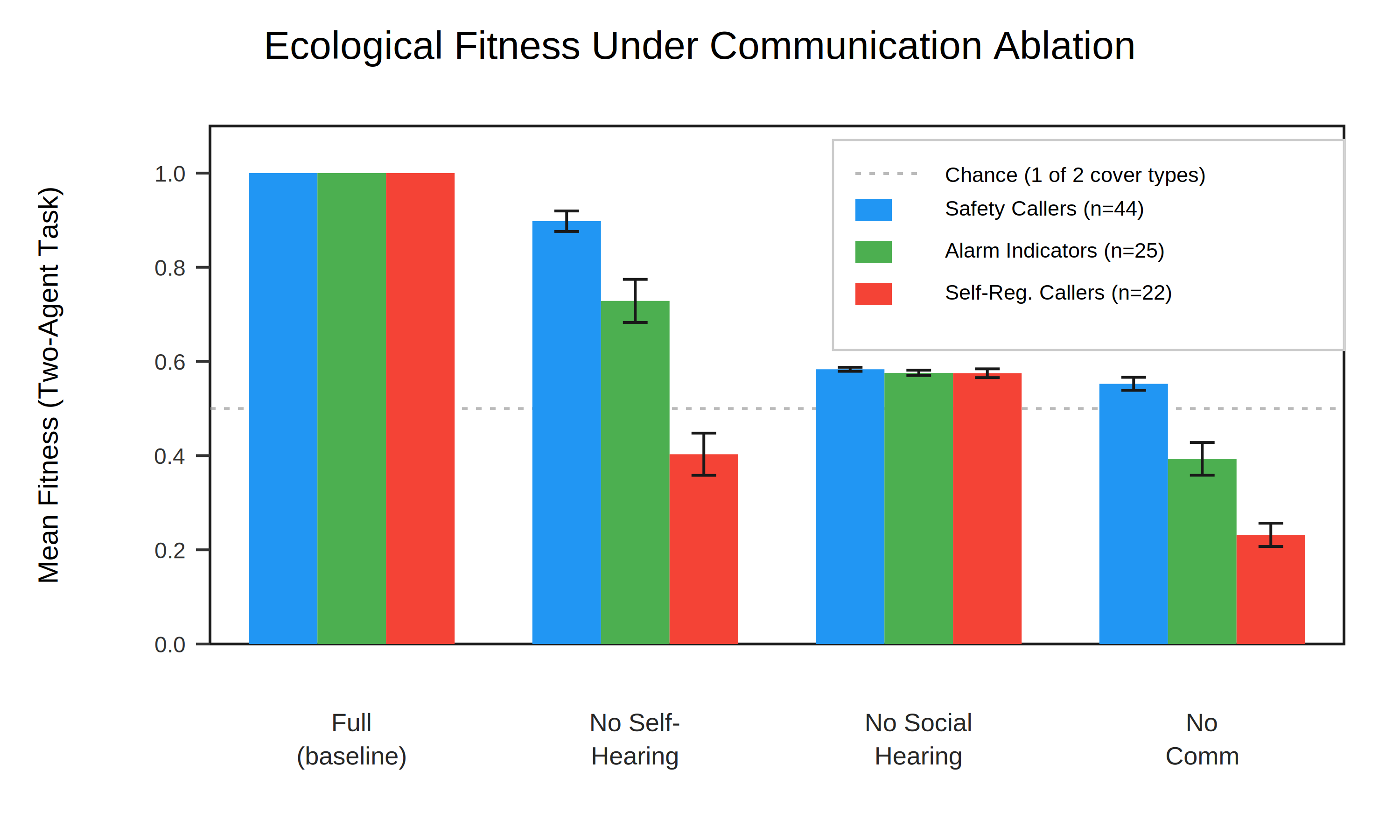}
    \caption{Ecological fitness under communication ablation by strategy. All agents achieve perfect fitness (1.0) under full communication. Removing self-hearing (NoSelf) selectively impairs self-regulatory callers, while safety callers and alarm indicators remain functional. Removing social hearing (NoSocial) affects all strategies equally. Error bars show standard error of the mean.}
    \label{fig:ecological}
\end{figure}

The results support the causal prediction (Figure~\ref{fig:ecological}, Table~\ref{tab:ecological}). Removing self-hearing has different effects depending on communication strategy ($p < 10^{-9}$). Safety callers maintain high fitness ($0.90$). Alarm indicators retain moderate fitness ($0.73$), suggesting they benefit from but do not depend on self-hearing. Self-regulatory callers suffer severe fitness loss ($0.40$; $p < 10^{-9}$ vs.\ safety callers). For these agents, communication helps sustain action: without hearing their own vocalization, they cannot maintain the behavioral response that vocalization supports. Removing social hearing reduces all three strategies equally to approximately 0.58 ($p = 0.52$), suggesting that the self-hearing effect is specific rather than a general communication effect.

\begin{table}[h]
\centering
\small
\begin{tabular}{lcccc}
\toprule
Strategy & $n$ & NoSelf & NoSoc. & NoCom. \\
\midrule
Safety call. & 44 & 0.90 & 0.58 & 0.55 \\
Alarm ind. & 25 & 0.73 & 0.58 & 0.39 \\
Self-reg. call. & 22 & 0.40 & 0.58 & 0.23 \\
\bottomrule
\end{tabular}
\caption{Mean ecological fitness (two-agent task) under communication ablation. All agents achieve 1.0 under Full communication. Self-regulatory callers lose the most fitness when self-hearing is removed, while safety callers and alarm indicators remain functional.}
\label{tab:ecological}
\end{table}

\textbf{Self-hearing dependency is graded, not binary.} Alarm indicators show intermediate NoSelf fitness (0.73), significantly below safety callers (0.90, $p < 0.01$) but well above self-regulatory callers (0.40, $p < 10^{-4}$). This gradient suggests that self-hearing dependency varies across agents that call during threat, with self-regulatory callers at the extreme end. Calling during ongoing escape creates the conditions for evolution to recruit the self-hearing channel, and the degree of reliance on that channel varies continuously.

\textbf{Removing both channels shows the dual role of communication.} For safety callers, removing social hearing and removing all communication produce nearly identical fitness drops (0.58 vs.\ 0.55), indicating that self-hearing contributes little. This fits the interpretation that safety calling is a consequence of behavior. For self-regulatory callers, removing all communication (0.23) is far worse than removing only self-hearing (0.40). This larger drop suggests that these agents use communication both for behavioral regulation (via self-hearing) \textit{and} to guide the partner (via social hearing). Removing one pathway impairs one function; removing both produces a severe fitness loss (0.23).

\section{Discussion}

\subsection{Indication and Self-Regulation as Distinct Evolved Functions}

Our results show that a single, minimal coordination task gives rise to three functionally distinct communication strategies, each best understood through a different lens. Alarm indication is broadly consistent with the standard information-transfer view: the signal correlates with an external state (predator presence) that the partner also needs to respond to. Safety calling is better understood as a state-dependent signal. When the agent reaches safe cover, its neural state shifts and it begins to call. The partner can use that call, but the caller does not need to hear it in order to produce it. Self-regulatory calling represents yet a different function: the vocalization is a component of the caller's own control loop, sustaining escape behavior through self-hearing.

The emergence of self-regulation is made possible by a modeling choice that is absent from prior minimal models of evolved communication: \textbf{self-hearing}. In natural vocal systems, organisms routinely hear their own vocalizations, yet computational models typically treat communication as a unidirectional channel from sender to receiver. By including self-hearing (allowing agents to receive their own communication output), we create the conditions under which evolution can discover that vocalization serves not only the receiver but also the caller.

The three strategies differ in the causal direction between action and communication. Safety calling operates in an \textit{act-then-communicate} mode: the agent reaches safety and then signals. Alarm indication operates in a \textit{communicate-and-act} mode: the agent vocalizes and escapes independently, with neither depending on the other. Self-regulatory calling operates in a \textit{communicate-to-act} mode: the vocalization sustains the escape behavior. The self-hearing ablation results support this interpretation. When self-hearing is removed, safety callers maintain high performance (ecological fitness 0.90), consistent with communication following action. Self-regulatory callers, by contrast, lose the ability to act appropriately (fitness 0.40), consistent with communication helping to drive action. This difference in causal ordering explains why self-hearing dependency is concentrated among agents that call during threat rather than being uniformly distributed. Alarm calling produces a signal during ongoing action, creating an opportunity for self-hearing to shape behavior. Safety calling does not create the same opportunity because the critical action has already been completed.

\subsection{Implications for Biological Alarm Calls}

Our results offer a new perspective on puzzling observations about natural alarm calls. If vocalization can serve a self-regulatory function, it would offer one explanation for why animals sometimes call in contexts where the communicative benefit to others appears limited \citep{sherman1985alarm, townsend2012flexible}: the call's function may include modulating the caller's own behavioral state, rather than exclusively transferring information to others. The tight temporal coupling between call production and behavioral response observed in many species \citep{owings1977snake, owings1998animal} is also consistent with a self-regulatory mechanism. These findings also connect to the long-standing debate between informational and influence perspectives on animal communication \citep{dawkins1978animal, rendall2009what, owren2010redefining}. In our model, self-regulatory calling is a case where the signal functions partly by influencing the \textit{signaler itself}. This possibility is not central to either information-transfer or receiver-influence accounts, but it is compatible with the assessment/management view \citep{owings1998animal}.

Although our model is deliberately minimal, it generates specific testable predictions. First, species or individuals that alarm call during active escape (analogous to our self-regulatory callers) should be more disrupted by reductions in self-hearing than those that call from positions of safety. Sentinel species such as meerkats, which characteristically move to an elevated lookout position before vocalizing \citep{clutton1999selfish, manser2014vocal}, may resemble safety callers in that calling is linked to a state of relative safety. Species or individuals that call during active escape may resemble self-regulatory callers. Second, if self-regulatory callers depend on hearing their own vocalizations, acoustic noise masking should impair escape performance more than it impairs the performance of safety callers. The Lombard effect, the tendency to increase vocal amplitude in noisy conditions \citep{zollinger2011lombard}, may reflect not only a communicative adjustment but also a self-regulatory one, if maintaining audible self-hearing is itself behaviorally important. Third, audience-independent calling should be more common in contexts where calling co-occurs with active escape behavior, since the self-regulatory function does not require a receiver.

\subsection{Limitations}

As with all minimal models, our results are intended to demonstrate the existence of a phenomenon rather than to reproduce its full biological complexity. Several limitations bound the scope of our conclusions.

Our 5-neuron CTRNNs are minimal; whether self-regulation emerges in larger networks where internal memory is less constrained is an open question. The self-hearing channel may function as an external memory aid, providing a persistent signal that compensates for limited internal state capacity in small networks. More complex controllers with stronger recurrent memory may rely on self-hearing less. The agents were evolved as clonal pairs; robustness to heterogeneous pairing remains untested. The binary communication channel cannot capture the graded spectral information available in biological vocalizations.

Our model's one-timestep self-feedback also differs from biological self-hearing in delay and signal fidelity: acoustic propagation, neural processing latency, and spectral transformation are not modeled. Whether self-regulatory communication remains viable under more realistic feedback delays is an open question. Additionally, many species partially attenuate neural responses to self-generated sounds via corollary discharge \citep{eliades2008neural}; whether self-hearing remains available to influence behavior during alarm calling is an empirical question that our model helps motivate. Our strategy boundaries are based on thresholds, though the main strategy ordering persists under threshold variation and the graded pattern of dependency.

\section{Conclusion}

We presented a systematic analysis of 112 perfect-fitness CTRNN agents, selected from over 2{,}000 evolutionary runs, solving a minimal predator avoidance task in which communication is necessary for robust performance. By including self-hearing, a natural feature of biological vocalization that is absent from prior minimal models, we find that communication can evolve to serve qualitatively different functions within a single task.

Three dominant strategies emerge, accounting for 81\% of agents: safety callers (39\%) that signal from safe cover, alarm indicators (22\%) that signal threat without self-hearing dependency, and self-regulatory callers (20\%) that depend on hearing their own vocalization to maintain escape behavior. Self-hearing dependency is more common among agents that call during an active threat than among agents that call only after reaching safe cover (47\% vs.\ 10\%), reflecting the causal direction of each strategy: safety callers act first and then communicate, while self-regulatory callers communicate in order to act.

These results demonstrate that even minimal neural architectures discover multiple solutions to coordination problems, including solutions where vocalization serves the caller as much as the receiver. This suggests, by analogy, that the function of alarm calls in natural systems may extend beyond information transfer to include self-regulation, a possibility that the testable predictions outlined above may help evaluate in empirical work on animal communication.

\section*{Acknowledgments}
This work used the Big Red 200 supercomputer at Indiana University, supported by
Lilly Endowment, Inc., through its support for the Indiana University Pervasive
Technology Institute. AI tools were used to assist with editing the manuscript and
assisting with code development. Multiple independent AI systems were employed, with
outputs cross-referenced to identify inconsistencies. The author reviewed all
AI-generated and AI-edited content, taking full responsibility for the final
work. All code was further validated through systematic testing and
visualization.

\footnotesize
\bibliography{references}
\bibliographystyle{apalike}

\end{document}